\documentclass[12pt]{iopart}

\usepackage{iopams}
\usepackage{graphicx}
\usepackage{subfigure}
\usepackage{color}
\usepackage{latexsym}
\usepackage{array}
\usepackage{bbding}
\begin{document}

\title[Active particles in geometrically confined viscoelastic fluids]{Active particles in geometrically confined viscoelastic fluids}

\author{N Narinder$^{1}$, Juan Ruben Gomez-Solano$^{2}$, and Clemens Bechinger$^{1}$}

\address{$^1$Fachbereich Physik, Universit\"at Konstanz, Konstanz, D-78464, Germany}
\address{$^2$Instituto de F\'isica, Universidad Nacional Aut\'onoma de M\'exico, 04510, Mexico City, Mexico}

\ead{clemens.bechinger@uni-konstanz.de}

\vspace{10pt}
\begin{indented}
\item[]May 2019
\end{indented}

\begin{abstract}
We experimentally study the dynamics of active particles (APs) in a viscoelastic fluid under various geometrical constraints such as flat walls, spherical obstacles and cylindrical cavities. We observe that the main effect of the confined viscoelastic fluid is to induce an effective repulsion on the APs when moving close to a rigid surface, which depends on the incident angle, the surface curvature and the particle activity. Additionally, the geometrical confinement imposes an asymmetry to their movement, which leads to strong hydrodynamic torques, thus resulting in detention times on the wall surface orders of magnitude shorter than suggested by thermal diffusion. We show that such viscoelasticity-mediated interactions have striking consequences on the behavior of multi-AP systems strongly confined in a circular pore. In particular, these systems exhibit a transition from liquid-like behavior to a highly ordered state upon increasing their activity. A further increase in activity melts the order, thus leading to a re-entrant liquid-like behavior. 
\end{abstract}

%
%
%
%
%

\noindent{\it Keywords}: active matter, self-propelled particles, viscoelastic fluids, collective motion, geometrical confinement, hydrodynamic interactions

\section{Introduction}

The natural habitat of microorganisms is often rather complex not only in geometrical  aspects~\cite{bechinger2016} but  also because the surrounding fluid environment, due to presence of colloids and macromolecules, exhibits viscoelastic behavior~\cite{patteson2016}. Thus, depending on the characteristic deformation rate, such fluid environment can display the characteristic behavior of either an elastic solid or a viscous liquid \cite{lauga2009}. In recent years, considerable work has been devoted to understand the role of the fluid properties on the self-propulsion of motile microorganisms. By now, there are ample examples demonstrating that their swimming motion depends strongly on the rheological properties of the fluid. \emph{E. coli}, for instance, swims in a completely different manner in polymeric solutions where it exhibits enhanced translational dynamics and reduced orientational ones \cite{patteson2015}; cervical mucus, a highly viscoelastic medium, alters the behavior of sperm cells by varying the wavelengths and amplitudes of their tails \cite{dillon2006};  sperms swim randomly with no alignment between their neighbours while moving in Newtonian liquids, however, viscoelastic fluid stimulates cell-cell alignment and thereby promotes their collective swimming \cite{tung2017,ishimoto2018}. Furthermore, the presence of confining walls and patterned surfaces, can significantly affect the motion of microswimmers compared to that in the bulk~\cite{zoettl2016}. For instance, steric, phoretic and hydrodynamic interactions can modify the motility of individual sperm cells navigating in microchannels ~\cite{rode2019}, as well as the collective behavior of crowded bacterial colonies under confinement~\cite{lushi2014}. Investigating such processes is also of major importance for the understanding of intracellular motility, as many organelles have to move  through highly confined viscoelastic media within cells~\cite{chaigne2016}.

Recently, these out-of-equilibrium systems have attracted tremendous interest in various scientific communities to design their artificial counterparts which mimic the self-propulsion of natural microswimmers \cite{niu2018}. This can be achieved using, e.g. acoustic \cite{wang2012}, magnetic~\cite{baraban2013} and optical fields \cite{dai2016}. Understanding the behavior of these artificial agents in realistic complex environments is of great significance \cite{xiao2019}, since they can be employed in many biomedical applications such as  drug delivery \cite{patra2013}, cargo transport \cite{Schmidt2012,Studart2018}, bio-sensing \cite{Stevens2014} etc. Besides, an AP in a viscoelastic fluid represents an example of a random walker in a non-equilibrium thermal bath, being of fundamental relevance for non-equilibrium statistical physics \cite{berner2018}. Despite holding such immense potential, theoretical studies involving the dynamics of self-propelled particles in complex fluids are rather scarce \cite{zhu2012,decorato2015,datt2015,datt2017,elfring2017,natale2017,datt2018,aragones2018,puljiz2019,du2019}. Experiments dealing with artificial microswimmers in viscoelastic fluids, demonstrate remarkable differences compared to entirely viscous environments \cite{gomezsolano2016,narinder2018}. For instance, instead of being predominantly controlled by thermal diffusion, as observed in Newtonian liquids \cite{howse2007,gomezsolano2017,chaterjee2018}, in viscoelastic fluids the particle orientation is strongly subjected to the slow response of the surroundings. This results in a transition of the orientational particle dynamics from enhanced rotational diffusion to persistent circular motion when increasing their propulsion speed \cite{narinder2018}. In addition, contrary to Newtonian fluids where APs accumulate near walls \cite{volpe2011}, get trapped by spherical colloids~\cite{takagi2014,brown2016}, and can be guided along topographic pathways \cite{simmchen2016,wykes2017} experiments under such spatial conditions in viscoelastic fluids have not been realized so far.

Here, as a natural extension towards more realistic conditions, we study the dynamics of synthetic APs in viscoelastic fluids under simple geometric constraints such as flat walls, fixed obstacles, and circular confinements. We find numerous novel features in the dynamics of active colloids moving in such complex environments, which can be interpreted in terms of effective interactions mediated by the viscoelastic surroundings with no counterpart in Newtonian fluids. We also repeated the experiments under similar conditions in a Newtonian fluid, where we show in a straightforward manner that the rotational behavior of APs remains largely unaffected by their interaction with solid surfaces, consistent with indirect experimental observations previously reported \cite{volpe2011}.

\section{Experimental details}

Active particles are made from silica spheres with diameter $2a = 7.82$ $\mu$m which are half-coated with a 30 nm carbon layer.  As solvent, we use a critical binary mixture of water and propylene glycol \textit{n}-propyl ether (PnP) which behaves as a Newtonian fluid with viscosity $\eta=0.004$ Pa s at temperature $T = 298$~K and exhibits a lower critical temperature at $T_c = 305$~K~\cite{bauduin2004}. At the critical composition (0.4 mass fraction of PnP) and below $T_c$, the binary mixture is homogeneous, whereas  it separates via spinodal decomposition upon increasing the temperature above $T_c$. In such a solvent, the rotational diffusion coefficient of our APs is $D_{r} = \frac{k_B T}{8\pi \eta a^3} = 6.8\times10^{-4}\,\mathrm{rad}^2\,\mathrm{s}^{-1}$. Active motion of the colloidal particles is achieved by laser illumination, which creates an anisotropic temperature distribution of the particle, which then leads to local demixing of the fluid, see \cite{samin2015,gomezsolano2017} for more details about the self-propulsion mechanism. Under the conditions considered here, the particle propels with the cap in the back. The propulsion velocity $v$ is controlled by the intensity $I$ of the illumination i.e. $v$ $\propto$ $I$. To render the fluid viscoelastic,  we added a small amount (0.05 wt.\%) of polyacrylamide (PAAm) to the mixture.  By means of passive microrheology (see \ref{app:microrheology}), we characterize the zero shear viscosity, $\eta_0= 0.210$~Pa~s, the effective solvent viscosity, $\eta_{\infty}= 0.039$~Pa~s, and the elastic modulus $G_0 = 9.9$~mPa of the resulting viscoelastic polymer solution. For such values, the long-time rotational diffusion coefficient in absence of light-induced self-propulsion is $D_{r} = \frac{k_B T}{8\pi \eta_0 a^3} = 1.3\times 10^{-5}$rad$^2\mathrm{s}^{-1}$, while the effective stiffness of the elastic forces exerted by the surrounding polymers is $\kappa = 6\pi a G_0 =  0.73 \, \mathrm{pN}\,\mu\mathrm{m}^{-1}$, which is comparable to those achieved by optical tweezers~\cite{gomezsolano2015}. Walls and circular confinements are fabricated from the photoresist SU-8 on glass slides by standard procedures of photolithography. The two-dimensional (2D) coordinates $(x,y)$ of the center of the particles and their in-plane orientation $\textrm{\textbf{n}}=(\cos\theta, \sin\theta)$, defined as a unit vector pointing from its capped to the uncapped hemisphere [see Figs. \ref{fgr:flatwall}(a) and \ref{fgr:flatwall}(b)], are obtained from video images using standard particle tracking algorithms \cite{gomezsolano2017}.

\section{Active particles near a flat wall}
\label{sec:activeflatwall}
\begin{figure}
\centering
  \includegraphics[scale=0.275]{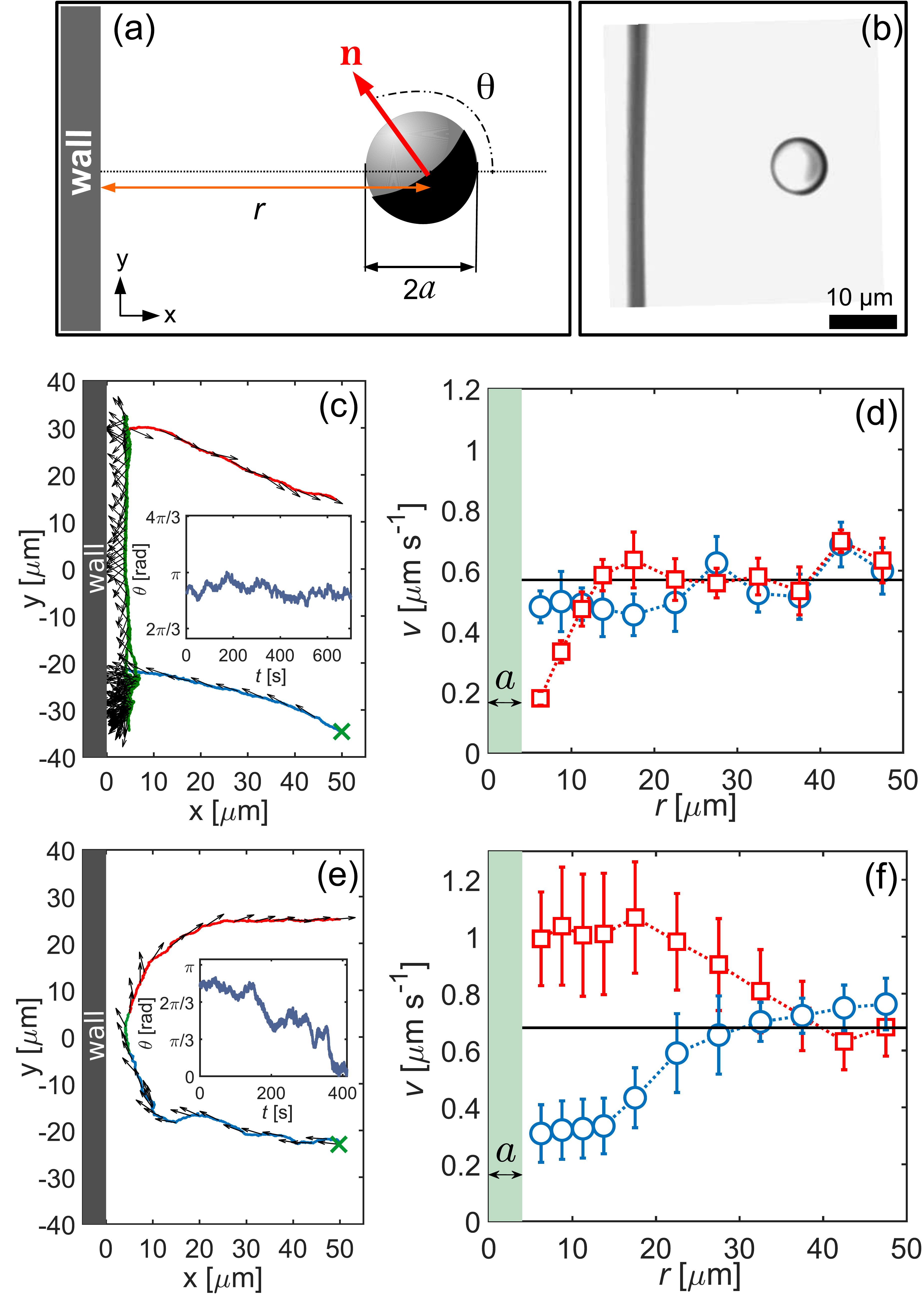}
  \caption{(a) Sketch of Janus particle of diameter $2a$ in presence of a flat wall. The particle orientation and its distance from the flat wall are denoted by $\mathbf{n}$ and $r$ respectively. (b) Snapshot (top-view) of an AP encountering a rigid flat wall. (c) Typical trajectory of a self-propelled particle of diameter $2a=7.82$ $\mu$m in the Newtonian fluid interacting with a rigid flat wall. The initial position of the particle is marked with the symbol $(\times)$. The arrows represent the instantaneous particle orientation. Inset: Time evolution of the angle $\theta$. (d) Particle speed $v$ as a function of the distance to the wall, $r$, in the Newtonian fluid during the approach ($\circ$) and during the departure ($\square$). (e) Behavior of the active particle approaching perpendicularly to a flat wall and then departing from it in the viscoelastic fluid. The symbol $(\times)$ marks the starting position of the particle trajectory. Inset: Time evolution of the angle $\theta$. (f) Particle speed $v$ as a function of the distance to the wall $r$ in the viscoelastic fluid during the approach (o) and during the departure ($\square$). The horizontal solid line in both (d) and (f) represents the particle speed very far from the wall.}
  \label{fgr:flatwall}
\end{figure}

To understand how the interaction of spherical APs with a solid surface is modified by the presence of a viscoelastic fluid (compared to a Newtonian fluid), we start by discussing the simplest geometry, i.e. a flat  wall. An AP interacting with a flat wall is sketched in Fig.~\ref{fgr:flatwall}(a), where $r$ denotes the distance of the particle to the wall.  An experimental snapshot of the top view of an AP interacting with a flat wall is shown in Fig.~\ref{fgr:flatwall}(b). Figure ~\ref{fgr:flatwall}(c) shows the typical trajectory of an AP moving in a Newtonian water-PnP fluid. Once the particle arrives at the surface, it gets trapped as long as the component of its orientation vector $\mathbf{n}$ perpendicular to the surface tangent remains finite, i.e. $\cos \theta < 0)$ [see arrows in Fig. ~\ref{fgr:flatwall}(c)]. On the other hand, a finite component of $\mathbf{n}$ parallel to the wall, i.e. $\sin \theta \neq 0$, leads to a lateral displacement of the AP. The particle confinement near the wall persists until $\mathbf{n}$ points away from the surface, i.e. when $\cos \theta > 0$. Since the orientational motion of APs in viscous fluids is almost entirely determined by their rotational diffusion time $1/D_{r} \approx 25$~min, this sets the typical time scale for particle trapping at walls under such given conditions.

From the particle trajectory, we obtain the AP velocity $v$ as a function of its distance $r$ to the wall. The results (averaged over 20 trajectories) are plotted in Fig.~\ref{fgr:flatwall}(d), during the approach (blue) and departure (red) of an AP. We find that $v$ remains rather constant and very close to the speed far away from the wall. A reduction in $v$ is observed only during the detachment from the wall, which is possibly due to phoretic interactions induced by the concentration gradient adjacent to the capped hemisphere when the particle cap faces the wall surface \cite{roy2018}. Therefore, in a Newtonian fluid the interaction between the AP and the wall surface is mainly mediated by steric forces and to a lesser extent by hydrodynamic effects.

A qualitatively different behavior is found in case of a viscoelastic fluid as shown in Fig.~\ref{fgr:flatwall}(e). In particular, the AP exhibits a pronounced systematic rotation of $\approx\pi$ rad in only $400$~s upon propelling towards the wall, as plotted in the inset of Fig.~\ref{fgr:flatwall}(e). This is in contrast with the rotational behavior in the Newtonian fluid, where the particle orientation is hardly affected by the solid surface over a similar time-scale, see inset of Fig.~\ref{fgr:flatwall}(c), here the AP undergoes a systematic rotation. Moreover, during approach the particle velocity $v$ significantly decreases even at distances $r\approx6a$ from the wall [shown in Fig.~\ref{fgr:flatwall}(f)]. More remarkable is the fact that the typical residence time of the AP on the wall surface is $\sim 1$~min, which is orders of magnitude below the rotational diffusion time  ($1/D_{r}\sim$ 21 h) of a passive particle in the viscoelastic fluid, after which the AP quickly rotates and leaves the wall surface. Right after the detachment, its speed $v$ is highly enhanced (approximately 2 times of its value far from the wall) and takes a large distance ($r\approx6a$) to recover to its value far from the wall.

We attribute the observed behavior to the accumulation and thereafter relaxation of the stress in the micro-structure of the viscoelastic fluid close to the boundary~\cite{puljiz2019}. As the particle approaches the wall by self-propulsion, it strains the polymers in the fluid between its own surface and the wall. Moving against the confined fluid strongly opposes its motion which leads to a dramatic decrease in its speed. Moreover, the wall surface also induces an asymmetry to its motion, results in a hydrodynamic torque which makes the particle quickly rotate and depart from the wall~\cite{becker1996,feng1996}. When a random change in the particle orientation allows the AP orientation to point away from the wall leads to a sudden release of the accumulated stress, thus resulting in a significant enhancement of its instantaneous velocity. Indeed, we check that the strength of this effect depends on the viscoelasticity of the fluid. The role of the fluid elasticity with respect to viscous effects can be quantified by the ratio $\frac{G_0 \tau}{\eta_0} = 1 - \frac{\eta_{\infty}}{\eta_0}$, where $\eta_{\infty}$  is the effective solvent viscosity and $\tau = \frac{\eta_0 -  \eta_{\infty}}{G_0}$ is the polymer relaxation time.  For instance, for a Newtonian fluid, $1 - \frac{\eta_{\infty}}{\eta_0} = 0$, whereas for a perfect Maxwell material $1 - \frac{\eta_{\infty}}{\eta_0} = 1$. For the data presented here at 0.05 wt.\% PAAm, $1 - \frac{\eta_{\infty}}{\eta_0} = 0.81$. In \ref{app:viscoelasticity} we show that, at 0.03 wt.\%, for which $1 - \frac{\eta_{\infty}}{\eta_0} = 0.26$, the viscoelasticity-mediated effect of the flat wall on the particle motion is less pronounced than at 0.05 wt.\% PAAm.

\section{Active particles near a spherical obstacle}
\begin{figure}
	\centering
	\includegraphics[scale=0.6]{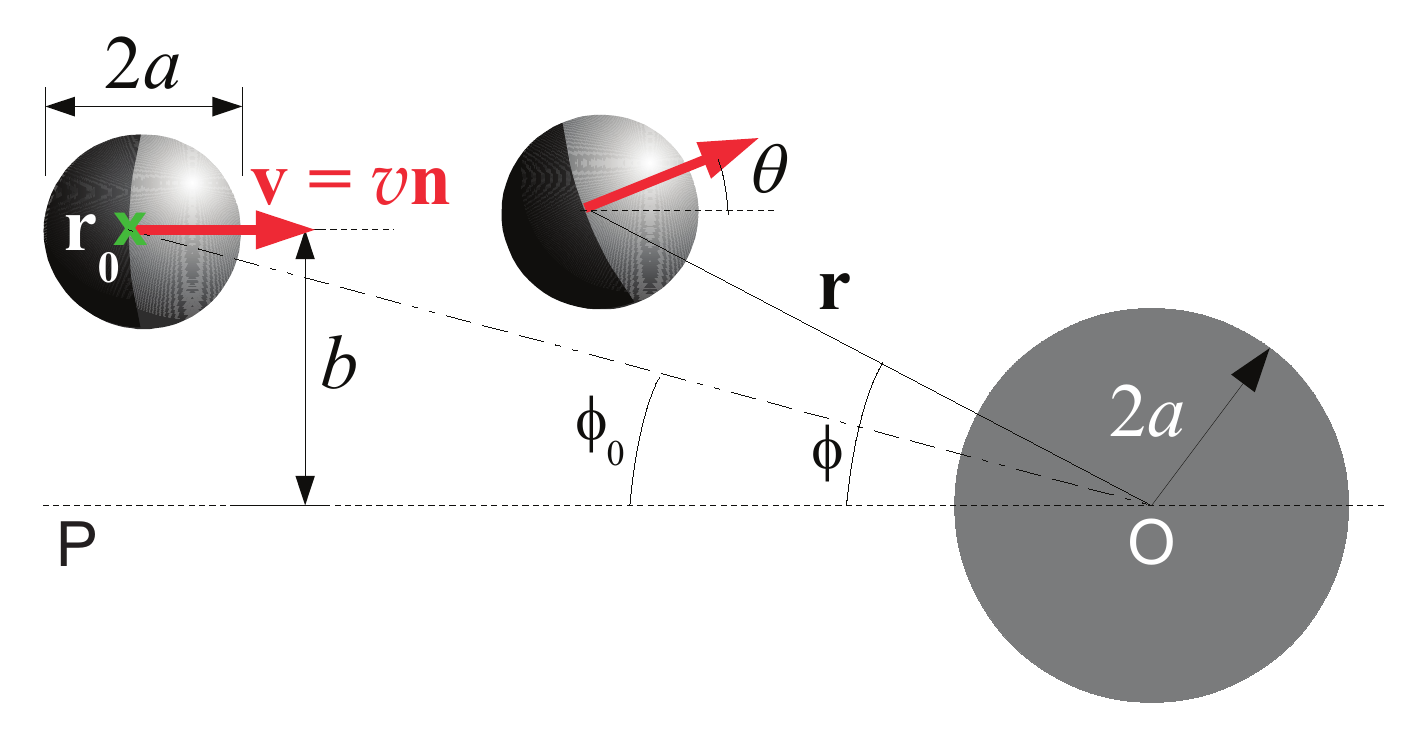}
	\caption{Sketch of the encounter of an active particle (radius $a$, orientation $\mathbf{n}$) with an immobile spherical obstacle (radius $2a$). The origin O of the coordinate system is located at the center of the obstacle, whereas P represents the polar axis. Here, $\mathbf{r} = (r,\phi)$ denotes the position of the active particle relative to O, where $r = |\mathbf{r}|$ and $\phi$ is the polar angle between $\mathbf{r}$ and P. The initial position of the active particle ($\times$) is $\mathbf{r}_0 = (r_0,\phi_0)$. The impact parameter, i.e. the distance of closest approach to O in absence of the obstacle, is denoted as $b$.}
	\label{fgr:obstaclesketch}
\end{figure}

Next, we study the interaction of the self-propelled particle with a spherical obstacle (radius $2a$), which remains immobile in the laboratory frame of reference. The coordinates we use to characterize this process are sketched in Fig.~\ref{fgr:obstaclesketch}, where the center of the obstacle determines the origin O of the 2D polar coordinates $(r,\phi)$. The initial direction of motion and position $r_0$ define the polar axis P, the initial polar angle $\phi_0$ and the impact parameter $b$, such that $\cos \phi_0 = - \mathbf{r}_0 \cdot \mathbf{n} / |\mathbf{r}_0|$ and $b = r_0 \sin \phi_0$. For our experimental conditions, $r_0 \approx 15a$, while $b$ is chosen from 0 to several times $a$ in order to investigate distinct behaviors resulting from the encounter with the obstacle.\\

\begin{figure}
	\centering
	\includegraphics[scale=0.6]{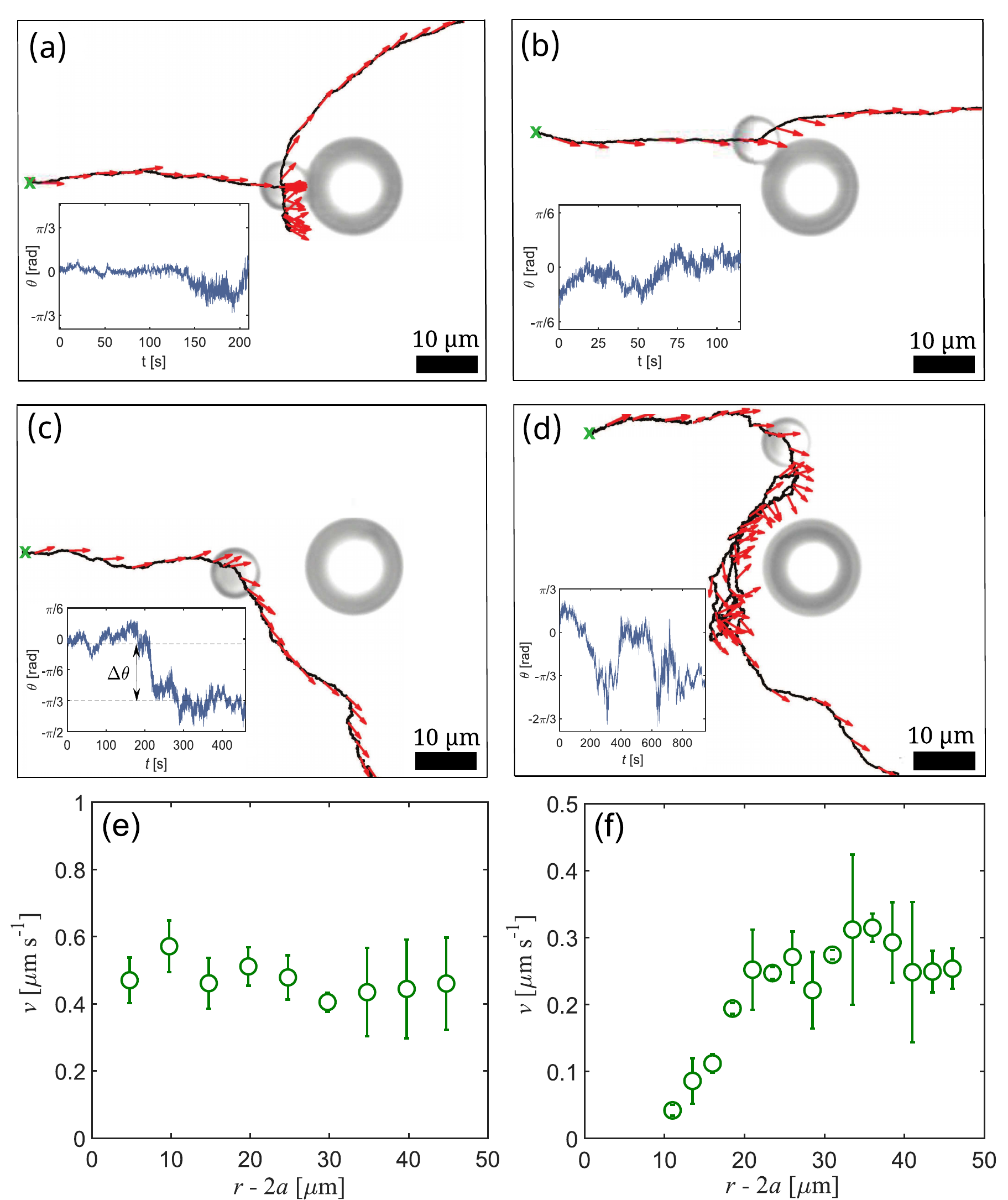}
	\caption{ Motion of a self-propelled spherical colloid of diameter $2a = 7.82$ $\mu$m close to a spherical fixed obstacle of radius $2a$ under different conditions: (a) Newtonian fluid, $b = 0$; (b) Newtonian fluid, $b = 2a$; (c) viscoelastic fluid, $b = 0$; (d) viscoelastic fluid, $b = 5a$.  The initial position of the particle, which defines the impact parameter $b$, is represented as a green ($\times$).  The black solid lines correspond to the trajectory of the self-propelled particle whereas the red arrows represent its instantaneous orientation $\mathbf{n}(t)$. Insets: time evolution of the angle $\theta$ of the corresponding trajectory in the main plot. (e) Speed of the particle approaching the spherical obstacle as a function of its distance to the obstacle surface $r-2a$, in Newtonian fluid for $b\approx 0$. (f) Speed of the particle approaching the spherical obstacle at $b\approx 0$ as a function of its distance to the surface of the obstacle in the viscoelastic fluid.}
	\label{fgr:obstacles}
\end{figure}

In  Figs.~\ref{fgr:obstacles}(a)-(d), we show that the motion of an AP in presence of an obstacle is strongly affected by the rheological properties of the surrounding fluid (Newtonian or viscoelastic) and depends on the impact parameter $b$. Figs.~\ref{fgr:obstacles}(a) and (b) show the particle trajectories for a Newtonian fluid, where the solid lines and the arrows represent the instantaneous particle positions and orientations, respectively.  In such a case, the rotational diffusion time ($1/D_{r} \approx 25$ min) is much larger than the typical time that the particle needs to arrive by self-propulsion $ (v \approx 0.46$ $\mu$ms$^{-1})$ at the distance of closest approach to obstacle surface ( in absence of the obstacle),  which is $r_{0} / v$ $\approx$ 2 min.  Consequently, the particle trajectory is expected to be rather straight with negligible changes in its initial orientation, as verified in Figs.~\ref{fgr:obstacles}(a)-(b). If the impact parameter (as defined in Fig.~\ref{fgr:obstaclesketch}) is much smaller than $a$, i.e. $b \approx 0$, the particle arrives with an orientation almost perpendicular to the obstacle surface,  and halts there, see Fig.~\ref{fgr:obstacles}(a). Then, it eventually escapes when its orientation is tangential to the obstacle surface, which occurs at a time scale $\lesssim$ $1/D_{r}$ over which $\mathbf{n}$ has a significant change by thermal diffusion.  With increasing $b$, the typical detention time at the obstacle, i.e. the time the microswimmer resides on its surface before completely leaving it~\cite{schaar2015}, decreases, as a smaller orientation variation is needed for the particle velocity to be tangential to it.  For example, in Fig.~\ref{fgr:obstacles}(b) we show that when $ b \approx 2a$, for which the particle arrives with an angle 0.84 rad relative to the tangent, the detention time is $\approx$ $4a/v =$ 1 min during which it slides along its surface and then moves forward. In all cases, if $b \le 3a$, the active particle is able to get in contact with the obstacle, i.e. the distance of the particle center to the obstacle surface is $r - 2a \approx a$,  at which hydrodynamic interactions might become important. Note that, for $b > 3a$, the obstacle has no significant influence on the self-propulsion of the AP. 

In Fig.~\ref{fgr:obstacles}(c)-(d), we plot the particle trajectories for two different values of the impact parameter $b$ in the viscoelastic fluid. When $b \ll a$, the particle is not able to arrive at the obstacle surface as shown in Fig.~\ref{fgr:obstacles}(c). Instead, its motion slows down and the particle is eventually deflected at a rather larger distance $r-2a \approx 4a$ from the obstacle surface. Thereafter its orientation $\mathbf{n}$ undergoes a systematic rotation, with an angular change of $\Delta \theta \approx \pi / 3$, as shown in the inset of Fig.~\ref{fgr:obstacles}(c). The distance of closest approach of the particle center to the obstacle surface is $r - 2a \approx 3a$. We point out that during this scattering process, even though the particle is still actively moving, $\mathbf{n}$ does not coincide with the direction of motion. This shows the presence of an effective interaction induced by the strained viscoelastic fluid, which opposes the AP motion. Such a viscoelasticity-mediated repulsive force, whose strength increases with decreasing $r - 2a$ according to our experimental observations, depends on the impact parameter $b$ and on the self-propulsion speed $v _0$ of the AP, i.e. its speed in absence of confinement. For $b = 0$ and $v_0 = 0.30~ \mu\mathrm{m}\, \mathrm{s}^{-1}$, we can estimate the strength $f_{VE}$ of the induced repulsive force at the distance of closest approach from the balance with the propulsive force:$f_{VE} = 6\pi \eta_0 a v_0 \approx 4.6 \, \mathrm{pN}$, which is comparable to the typical  forces exerted by optical tweezers~\cite{gomezsolano2015}. The orientational change of the particle during the deflection takes place during  $\sim 5$ min, which is orders of magnitude shorter than the rotational diffusion time  $D_{r}^{-1} \approx 21$~h of a passive particle of the same size embedded in the viscoelastic fluid. This effect suggests the emergence of a hydrodynamic torque exerted by the strained viscoelastic fluid on the particle due to the strong asymmetry created by the solid surface of the obstacle, similar to that observed close to a flat wall. The encounter with the obstacle becomes more intricate with increasing $b$. In Fig.~\ref{fgr:obstacles}(d), we illustrate the resulting motion when the impact parameter is $b = 5a$. Note that for the propulsion velocities considered here, at this distance the obstacle would not have any influence on the  AP motion in a Newtonian fluid. As a results, the AP would just pass by. Interestingly, in the viscoelastic fluid, the obstacle strongly affects the motion of the active particle even at a rather large distance $r - 2a \sim 3a$ from its surface. At this distance, a hydrodynamic torque is imparted by the fluid, thus reorienting the AP towards the obstacle. Thenceforth, the particle begins to hover around the obstacle, with no direct contact, i.e. the distance between the particle center and the obstacle surface remains $r - 2a > a$. During the hovering, which takes longer ($\approx$ 15 min) than the deflection shown in Fig.\ref{fgr:obstacles}(c), the particle exhibits large orientational fluctuations due to the interplay between its activity, rotational diffusion and the asymmetry-induced torque exerted by the viscoelastic fluid. Finally, when \textbf{n} points away from the tangent to the surface, the particle is able to escape.

Similar to the behavior close to a flat wall, we observe that, while in a Newtonian fluid the presence of the obstacle does not modify significantly the AP speed, the viscoelastic fluid slows down the AP motion at rather large distances $r - 2a$ from the obstacle surface. This is illustrated in Figs. \ref{fgr:obstacles}(e) and  \ref{fgr:obstacles}(f), respectively, for $b \approx 0$.  Here, we show that during the motion toward the obstacle in the viscoelastic fluid, the particle speed begins to decrease at $r - 2a \approx 4a$, getting almost fully arrested at $r - 2a \approx 3a$ during the deflection described in Fig. \ref{fgr:obstacles}(c).

\begin{figure}
	\centering
	\includegraphics[scale=0.35]{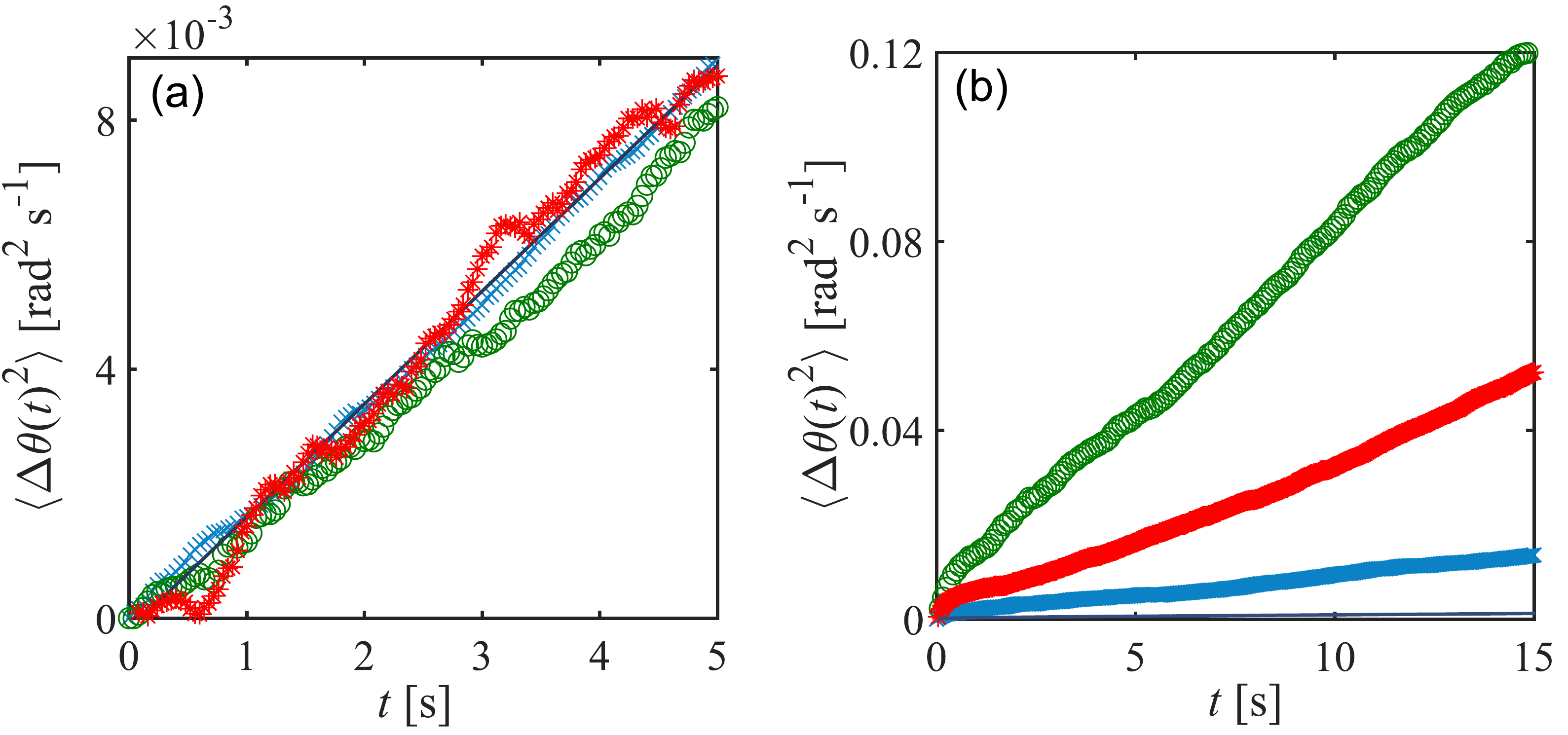}
	\caption{(a) Mean square angular displacement of an active particle moving in the Newtonian fluid far from the obstacle ($\times$), during the trapping ($b=0$) at the obstacle surface (o) and during the sliding($\ast$) which occurs at $b=2a$. (b) Mean square angular displacement of an active particle moving in the viscoelastic fluid far from the obstacle ($\times$),  during  the  deflection  by  the  obstacle  ($\ast$) at $b\approx0$ and during the hovering (o) at $b=5a$.  In both cases, the mean square displacements for a passive particle are represented as a dashed line.}
	\label{fgr:obstacleamsd}
\end{figure}

To better assess the effect of the hydrodynamic torque induced close to the obstacle surface, we compute the mean square displacement $\langle \Delta \theta(t)^2 \rangle$ of the angle $\theta$ between \textbf{n} and the polar  axis  P  [see  Fig.~\ref{fgr:obstacleamsd}] for the different processes previously described. In Fig.~\ref{fgr:obstacleamsd}(a) we show $\langle \Delta \theta(t)^2 \rangle$ for an active particle in the Newtonian fluid, computed far away from the obstacle, during the trapping shown in  Fig.~\ref{fgr:obstacles}(a) and during the sliding of Fig.~\ref{fgr:obstacles}(b).  In  all  cases,  within  our  experimental  resolution, $\langle \Delta \theta(t)^2 \rangle$ remains  unaffected  by  the  presence  of the  neighboring  solid  surface  of  the  obstacle and  exhibits diffusive behavior $\langle \Delta \theta(t)^2 \rangle = 2D_{r}t$.  In addition, the corresponding rotational diffusion coefficients $D_{r}$ is close  to  that  of  a  passive  colloid  [$v = 0$,  solid  line in Fig.~\ref{fgr:obstacleamsd}(a)], where the Stokes-Einstein relation holds, i.e. $D_{r} = k_B T/(8\pi \eta a^3) \approx 6.8\times10^{-4}\,\mathrm{rad}\,\mathrm{s}^{-1}$. From this agreement, we conclude that hydrodynamic and phoretic effects are only of vanishing importance for the APs rotational motion. This is in stark contrast with the behavior in the viscoelastic fluid, where we observe a strong influence of the approaching process to the obstacle, determined by the parameter $b$, as shown in Fig.~\ref{fgr:obstacleamsd}(b). Far away from the obstacle $(r_{0} \geq 10a)$, the mean square angular displacement of the active colloid is diffusive at sufficiently long time-scales, i.e. $\langle \Delta \theta(t)^2 \rangle = 2D_{r}t$,  with  an  effective  rotational  diffusion  coefficient higher than that for a passive colloid, as reported in Ref.~\cite{gomezsolano2016}, For  instance,  when  the  self-propelled  particle  moves  at $v \approx 0.10 \,\mu\mathrm{m}\,\mathrm{s}^{-1}$, $D_{r}$ is  approximately  50  times  higher than the value $D_{r} = k_B T/8\pi \eta_{0} a^3 \approx 1.3\times10^{-5}\,\mathrm{rad}\,\mathrm{s}^{-1}$ given by the Stokes-Einstein relation, see Fig.~\ref{fgr:obstacleamsd}(b). Remarkably, the close presence  of  the  obstacle  enhances even more the rotational diffusion of the active colloid. During the deflection, which occurs at $b \approx 0$, $D_{r}$ is approximately 200 higher than $D_{r} = k_B T/8\pi\eta_{0} a^3$, while during the hovering, $D_{r}$ is enhanced up to 450 times.

\begin{figure}
	\centering
	\includegraphics[scale=0.55]{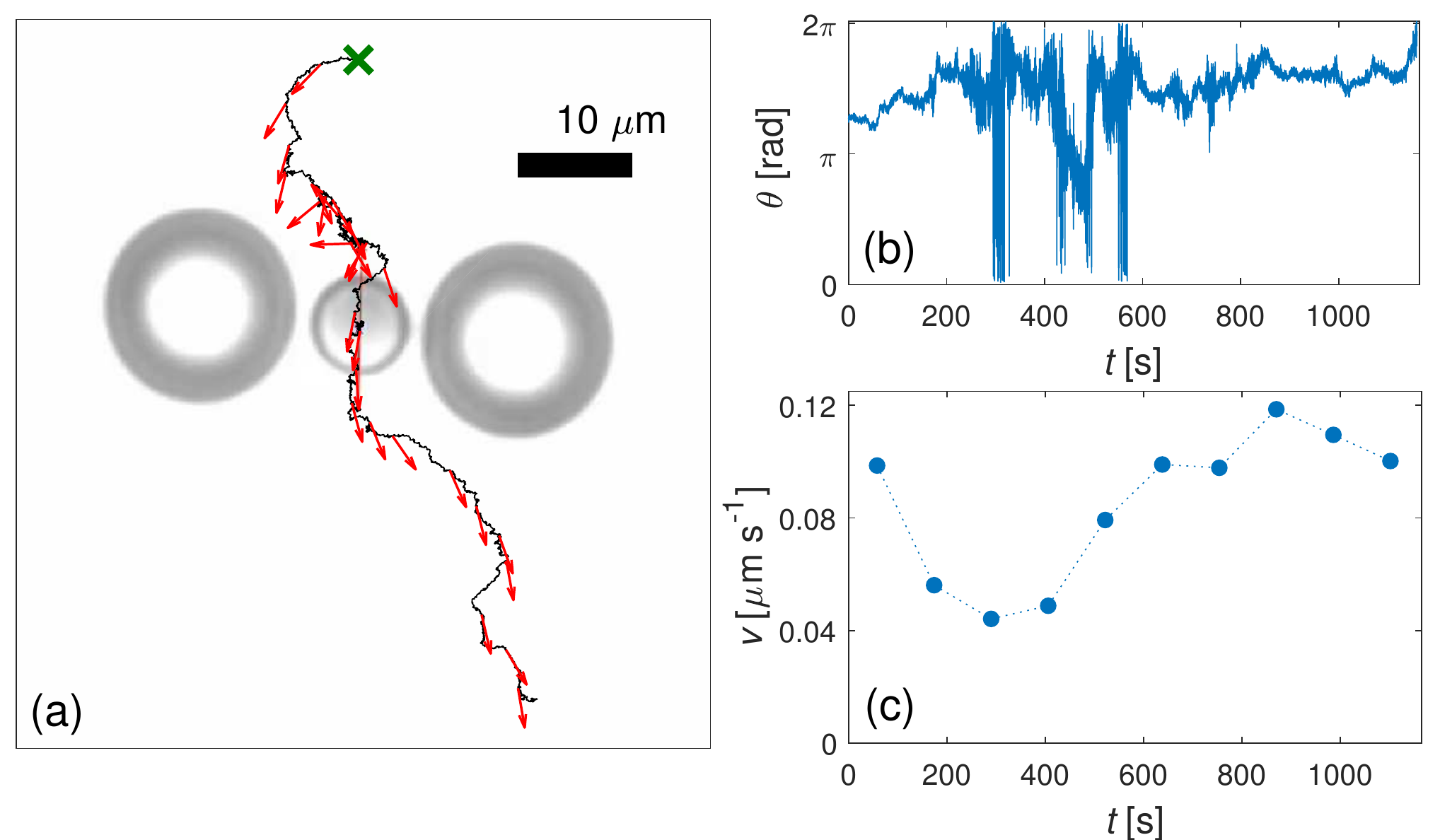}
	\caption{ (a) Trajectory (black solid line) and orientation (red arrows) of an AP getting close to two spherical obstacles in a viscoelastic fluid. The symbol ($\times$) represents the initial particle position. (b) Time evolution of the orientation angle $\theta$. (c) Time evolution of the AP speed.}
	\label{fgr:2obstacles}
\end{figure}

The aforementioned effective repulsion induced by a curved rigid surface permits local trapping and guidance of APs moving in viscoelastic fluids. For instance, in Fig. \ref{fgr:2obstacles}(a) we show the trajectory of an AP getting close to the gap between two immobile obstacles. During the approach, the AP orientation angle  $\theta$ displays large fluctuations, while its velocity decreases significantly over time, thus leading to a local trapping before entering the gap, as shown during the first 600 seconds in Figs. \ref{fgr:2obstacles}(b) and \ref{fgr:2obstacles}(c), respectively. The effective repulsive forces and torques induced by the two curves surfaces of the obstacles reorient the particle and lead to a guided motion within the gap, for which the fluctuations of $\theta$ decrease while the AP speed increases, as shown in in Figs. \ref{fgr:2obstacles}(b) and \ref{fgr:2obstacles}(c) for 600~s$< t <$~900~s. Finally, the particle is able to leave the gap for $t>$~900~s with a speed and angular motion very similar to those before the local trapping.

\section{Self-propelled particle in a circular confinement}\label{sec:activepore}
\begin{figure}
	\centering
	\includegraphics[scale=0.4]{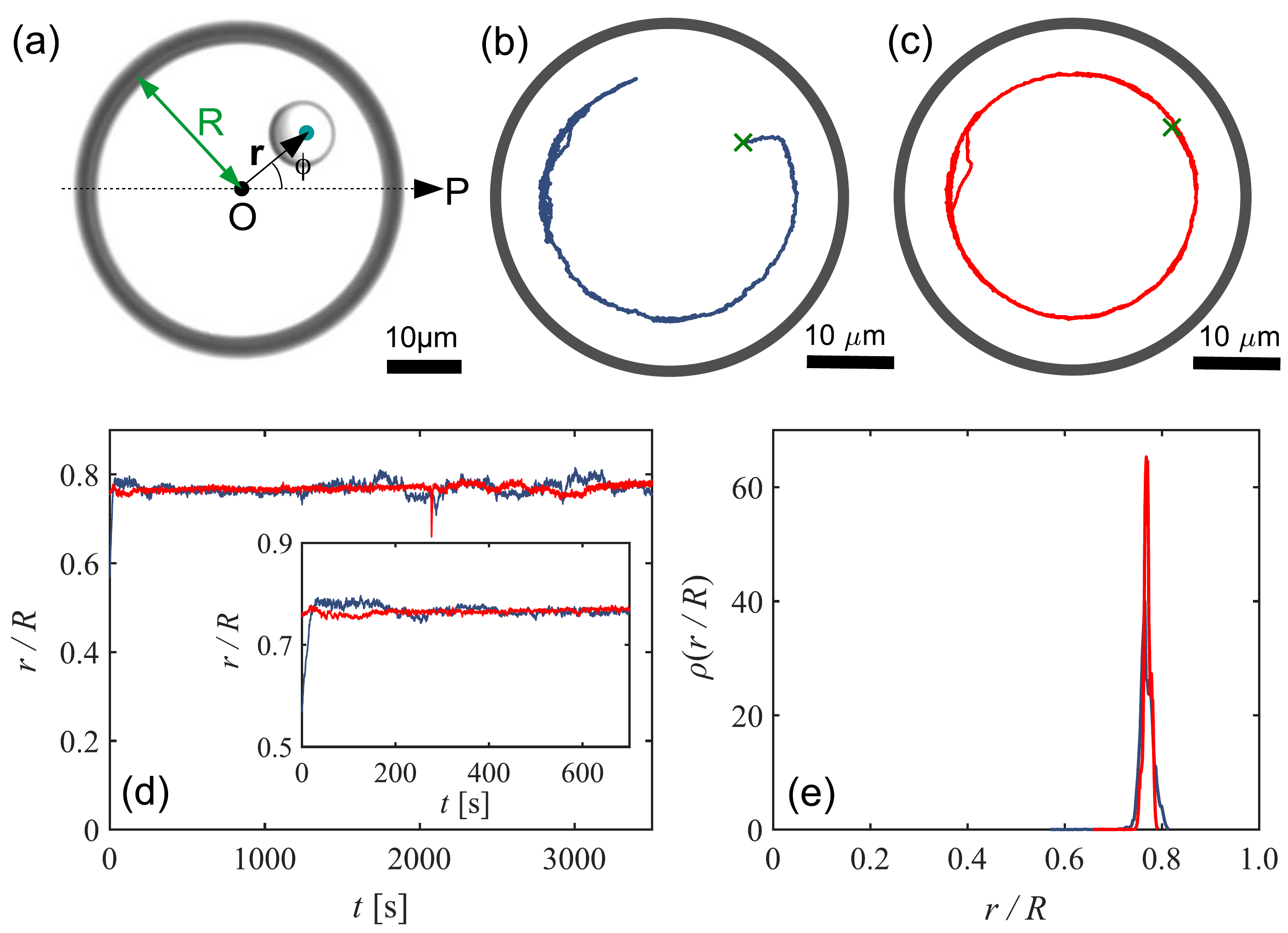}
	\caption{(a) Snapshot (top-view) of a Janus particle moving inside a circular confinement of radius $R=20~\mu\mathrm{m}$. The 2D motion of the AP is described in polar coordinates $\mathbf{r}=(r,\phi)$.  Trajectories of the center of mass of a self-propelled particle (radius $a = 3.9~\mu$m) moving in the Newtonian fluid under the circular confinement  at two different illumination intensities: (b) $I = 10~\mu\mathrm{W}/\mu\mathrm{m}^2$  and (c) $I = 15~\mu\mathrm{W}/\mu\mathrm{m}^2$. The starting position of the particle trajectory is labeled with $(\times)$. (d) Time evolution of the radial position of the trajectories. Inset: Expanded view of the main plot. (e) Normalized probability density function of the radial position of the particle.}
	\label{fgr:porenewtonianv1}
\end{figure}

We next study the behavior of the self-propelled particles in a cylindrical cavity of radius $R = 20$ $\mu\mathrm{m}$ in both a Newtonian and a viscoelastic fluid, which confines the 2D motion of the AP to the circular region $0 < r < R - a$. Here, we describe the motion by fixing the origin O of the polar coordinate system $(r,\phi)$ at the center of the confinement [Fig.~\ref{fgr:porenewtonianv1}(a)]. In Fig.~\ref{fgr:porenewtonianv1}(b) we show an exemplary trajectory of an AP (measured over 3600 s) moving in the Newtonian fluid under cylindrical confinement of diameter $2R=20~\mu$m at illumination intensity $I = 10~\mu\mathrm{W}/\mu\mathrm{m}^2$ which, in absence of the confinement, results in AP velocity $v_0\approx0.80~\mu\mathrm{m}$s$^{-1}$. Under such condition, the persistent length of the particle, $l_\mathrm{p} = v_0 /D_{r}\approx 1200~\mu$m, is much larger than the pore size i.e. $l_\mathrm{p} \gg 2R$. We observe that the AP remains on the pore wall surface and slides along the wall during the entire measurement time, reminiscent to the wall accumulation of microswimmers~\cite{jenselgeti2013,berke2008,volpe2011,sartori2018}. This is because its outward radial motion (relative to the pore center O) is stopped by the encounter with the rigid surface. Similar to the particle-flat wall interaction as mentioned in Sec. \ref{sec:activeflatwall}, the AP slides along the wall when its orientation has a non-zero component parallel to the surface tangent. Note that a full detachment  from  the wall is  only possible when the particle orientation has a non-zero radial component pointing toward the pore center. Since the orientational motion is dominated by rotational diffusion, as verified by the examples previously described, the probability to observe  such  events  is  very  small  under  our  experimental conditions. In addition, when detachment from the wall happens, as observed on the left side of Fig.~\ref{fgr:porenewtonianv1}(c), the particle quickly encounters the wall again, thus leading to a persistent radial trapping. In Fig.~\ref{fgr:porenewtonianv1}(d) we plot the time evolution of the radial component of the particle position. In the inset of Fig.~\ref{fgr:porenewtonianv1}(d) we show that apart from thermal fluctuations, such radial trajectories remain tightly  confined  around  the  value $r =  0.78 R$,  which is very close to the distance $0.8 R$ at which the particle and the wall are in contact.  We also find that the probability density function of the radial position, $\rho(r/R)$,  peaks sharply at $r = 0.78 R$ as shown in Fig.~\ref{fgr:porenewtonianv1}(e). The behavior of AP remain unaltered upon further increasing $I$, as demonstrated by the trajectory in Fig.~\ref{fgr:porenewtonianv1}(c) and red curve in Fig.~\ref{fgr:porenewtonianv1}(d) and (e). 

\begin{figure}
	\centering
	\includegraphics[scale=0.5]{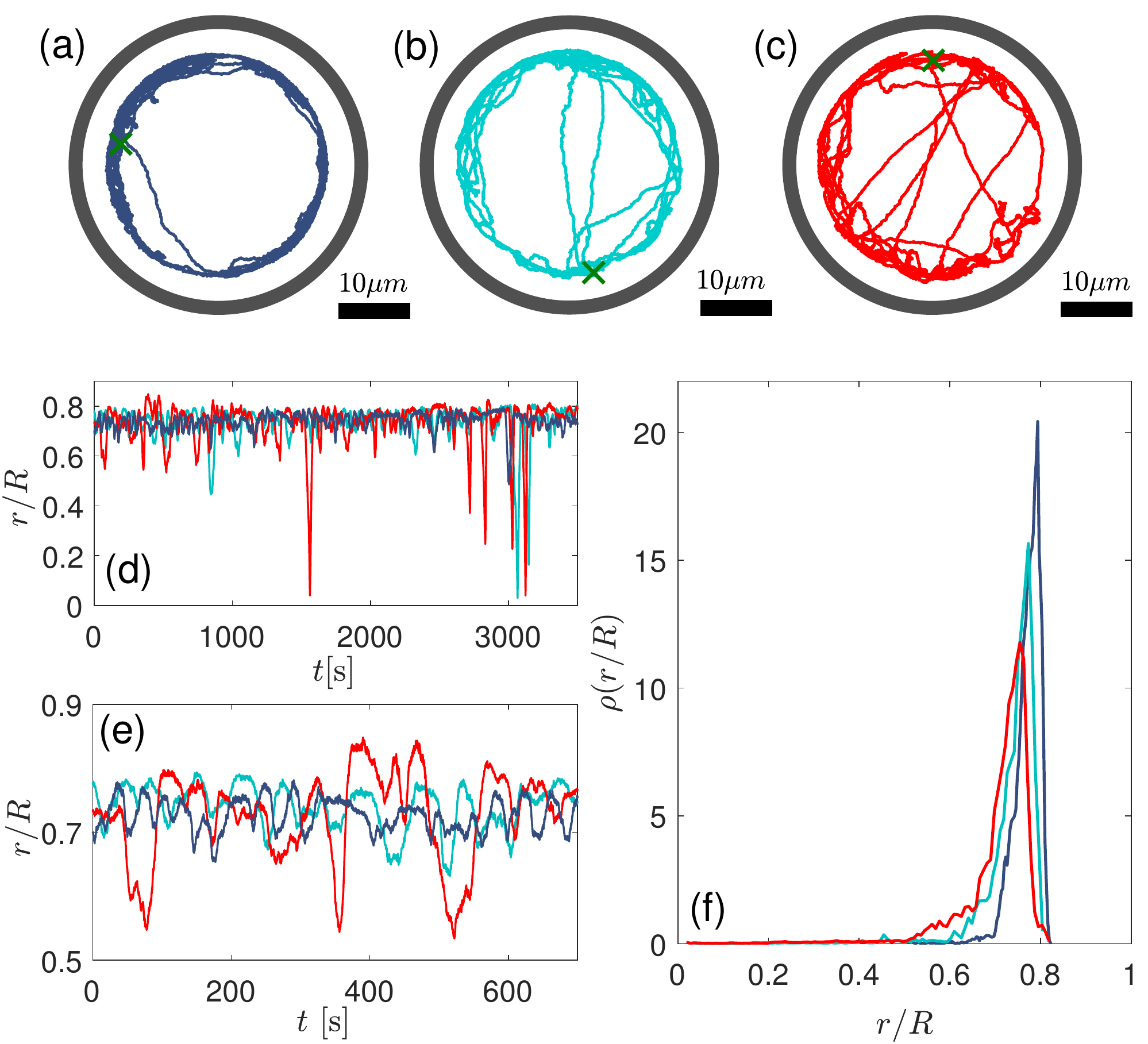}
	\caption{(a)-(c) Exemplary trajectories of the self-propelled particle of $2a = 7.82$ $\mu$m in the viscoelastic fluid at different illumination intensities $I$, (a) $I= 8~\mu\mathrm{W}/\mu\mathrm{m}^2$, (b) $I = 10~\mu\mathrm{W}/\mu\mathrm{m}^2$, and (c) $I = 12~\mu\mathrm{W}/\mu\mathrm{m}^2$ inside a circular pore of diameter $2R = 40~\mu$m. The $(\times)$ marks the initial position of the particle. (d) Time evolution of the radial position of the particle for each case (a)-(c) labelled with same color code as their trajectory. (e) Expanded view of the time evolution of the radial trajectories. (f) Normalized probability density function of the radial position of the particle.}
	\label{fgr:poreviscoelastic}
\end{figure}

APs display a qualitatively distinct behavior when moving in viscoelastic fluids under such circular confinement. This is illustrated in Figs.~\ref{fgr:poreviscoelastic}(a)-(c), where we plot the trajectories of a particle in the viscoelastic liquid inside a circular pore at different $I$, measured over 3600 s.  We find with increasing $I$, the particle is repelled more and more often away from the pore walls. Such an observation suggests that, similar to the behavior close to flat and a spherical surfaces, the viscoelastic fluid offers an activity-dependent elastic repulsion to the particle when moving close to the curved pore wall. In  Fig.~\ref{fgr:poreviscoelastic}(d), we plot the time evolution of the radial position $r$ of the particle normalized by the pore radius $R$. A closer look at such radial trajectories reveals the existence of intermittent fluctuations, whose magnitude grows with increasing $I$, as shown in Fig.~\ref{fgr:poreviscoelastic}(e). They originate from the interplay of the self-propulsion and the repulsive interactions induced by the accumulation and the release of elastic stress between the rigid wall and the AP. It should also be noted that such fluctuations are completely absent when the surrounding fluid is Newtonian as observed in Fig.~\ref{fgr:porenewtonianv1}(d). The normalized probability density function of the radial position of the particle, $\rho(r/R)$, is plotted in Figure~\ref{fgr:poreviscoelastic}(e) for each case shown in Figs.~\ref{fgr:poreviscoelastic}(a)-(c). Contrary to the behavior in the Newtonian fluid, here $\rho(r/R)$ depends strongly on the applied laser intensity. In particular, with an increase in $I$, the location of the peak of $\rho(r/R)$ shifts towards the center of the pore, while the distribution broadens and remains finite up to rather smaller values of $r/R$. This can be regarded as an activity-dependent attractive potential on the AP directed toward the pore center.

\section{Circularly confined multi-particle system}

\begin{figure}
\centering
 \includegraphics[scale=0.5]{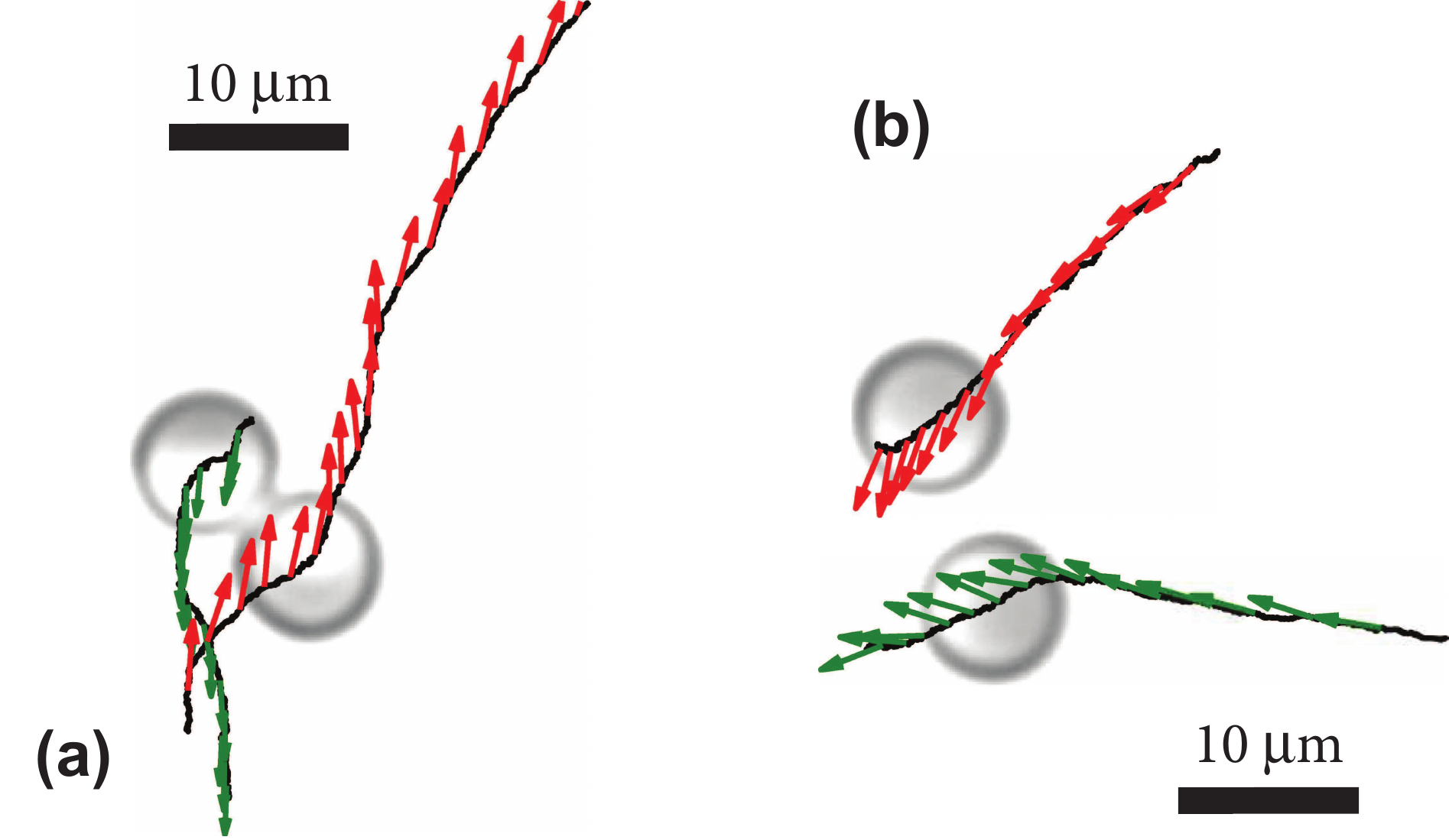}
 \caption{(a) Encounter of two self-propelled particles in the Newtonian fluid. (b) Encounter of two self-propelled particles in the viscoelastic fluid. The solid lines correspond to the particle trajectories whereas the arrows represent some of their instantaneous orientations.}
 \label{fgr:partpart}
\end{figure}

Confining several active  particles in a viscoelastic fluid turns the system very intricate because of two main reasons. The first is that they undergo an effective repulsion during the encounter of two or more particles, which is induced by the elastic response of the surrounding medium. Such particle-particle interactions are illustrated in Figs. \ref{fgr:partpart}(a) and \ref{fgr:partpart}(b) for the unconfined motion of two AP in the Newtonian and in the viscoelastic fluid, respectively. In Fig. \ref{fgr:partpart}(a) we show that, in the Newtonian fluid, the particles just get in contact, slide and move forward, where their relative orientations are governed by rotational diffusion. On the other hand, in the viscoelastic fluid, they undergo a mutual repulsion, which prevents them from touching each other, thus resulting in a deflection of their initial paths, as shown in Fig. \ref{fgr:partpart}(b). In addition, the fluid also exerts a torque on both particles, which makes them change their initial orientations more than expected by rotational diffusion during the approach. Note that such kind of behaviors are qualitatively similar to that observed for a AP moving near a stationary obstacle, but here the resulting trajectories can become much more complex due to the mobility of both particles.

\begin{figure*}
	\centering
	\includegraphics[scale=0.525]{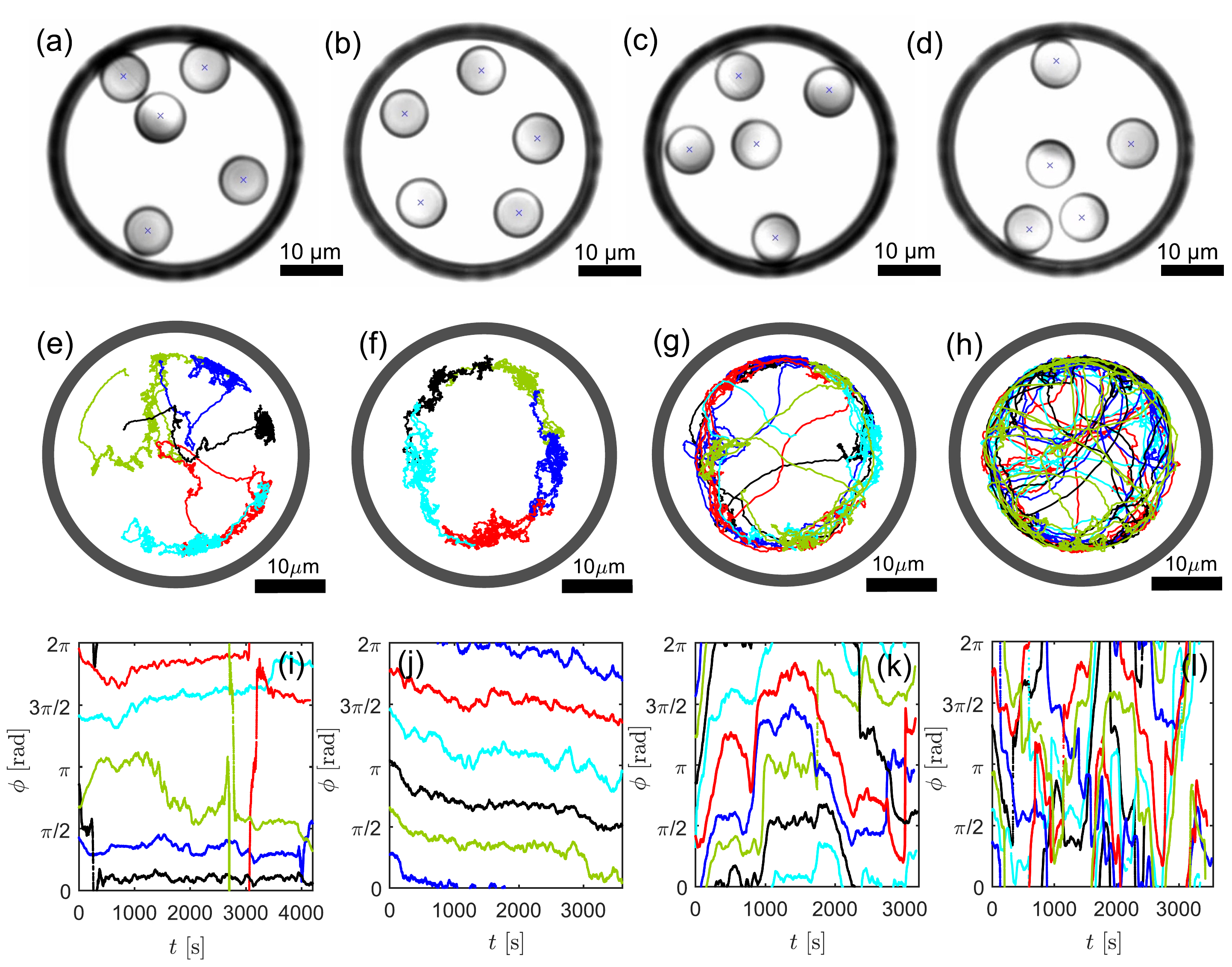}
	\caption{Snapshots (topview) of 5 active colloids ($2a=7.82~\mu$m) moving in a cylindrical confinement ($2R=40~\mu$m) in the viscoelastic fluid under various illumination intensities: (a) at $5~\mu$W$\mu$m$^{-2}$, (b) $6~\mu$W$\mu$m$^{-2}$, (c) $8~\mu$W$\mu$m$^{-2}$, (d) $10~\mu$W$\mu$m$^{-2}$. (e)-(h) Trajectories of the 5 self-propelled particles at different laser intensities which are as follows: (e) $5~\mu$W$\mu$m$^{-2}$, (f) $6~\mu$W$\mu$m$^{-2}$, (g) $8~\mu$W$\mu$m$^{-2}$, (h) $10~\mu$W$\mu$m$^{-2}$ measured over $\approx 3600$~s. (i)-(l) Time evolution of the polar angle $\phi$ for each of the 5 particles corresponding to the cases mentioned in (e)-(h) respectively.}
	\label{fgr:manyparticles}
\end{figure*}

Apart from the inter-particle effective interactions in the viscoelastic fluid, the rigid wall of a circular pore induces an effective force on a length-scale comparable to the pore radius $R$. Such a force attracts the APs to the pore center and its strength depends on the self-propulsion speed $v$, as described in the previous section. The combination of the inter-particle repulsive force and the attractive force results in intriguing collective motion of the multi-AP system. For instance, in Fig.~\ref{fgr:manyparticles}(e)-(h), we plot the trajectories of a system composed of $N = 5$ APs confined in a circular pore of $2R =  40~\mu$m under different values of $I$ over 3600 s. At low $I$, we find that particles explore the available space rather uniformly, even reaching the center of the pore during the measurement time, as depicted in Fig.~\ref{fgr:manyparticles}(e). Counter-intuitively, increasing $I$ slows down the collective dynamics of the active suspension. In particular, upon increasing $I$ to a certain value (6~$\mu$W$\mu$m$^{-2}$), their motion is highly localized both radially and polarly with respect to the pore center, see Fig.~\ref{fgr:manyparticles}(f). In this crystalline-like state, particles self-organize in a highly ordered fashion which obeys a rotational symmetry of order $N$ and such an order persists over the entire measurement. Increasing the activity further dissolves the order and a re-entrant liquid-like behavior is restored as evidently observed in Fig.~\ref{fgr:manyparticles}(g) and (h). 
We therefore uncover a transition from liquid-like behavior to a state of ordered arrangement and then to a re-entrant liquid-like behavior upon increasing the activity in strongly confined multi-particle system. Furthermore, in Figs.~\ref{fgr:manyparticles}(i)-\ref{fgr:manyparticles}(l) we show that the motion of all particles composing the confined system is strongly correlated, regardless of the value $I$. A variation in the angular position $\phi$ of a single particle is not independent from the rest, but it propagates through the entire system, giving rise to a collective motion.

\begin{figure*}
	\centering
	\includegraphics[scale=0.675]{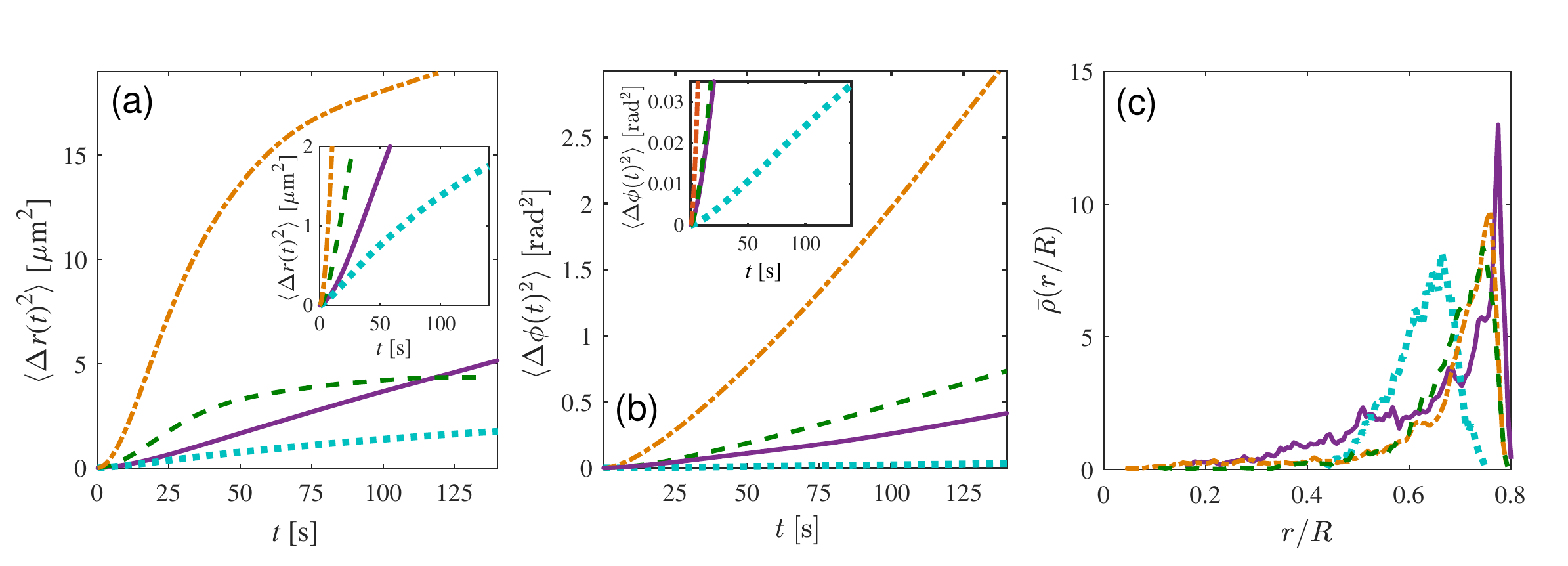}
	\caption{(a) Mean-square displacement of the radial particle position, $r$, for different intensities $I$: 5 $\mu$W$\mu$m$^{-2}$ (solid violet line), 6 $\mu$W$\mu$m$^{-2}$ (dotted turquoise line), 8 $\mu$W$\mu$m$^{-2}$ (dashed green line), and 10 $\mu$W$\mu$m$^{-2}$ (dotted-dashed orange line). Inset: expanded view around the order state at $I_c = 6\mu$W$\mu$m$^{-2}$ (see main text). (b) Mean-square displacement of the corresponding polar angle $\phi$, for different intensities $I$. Same linestyle and colorcode as in \ref{fgr:transition}(a). Inset: expanded view around the order state. (c) Radial probability density function at different intensities $I$. Same linestyle and colorcode as in \ref{fgr:transition}(a).}
	\label{fgr:transition}
\end{figure*}

What is the origin of such an unexpected transition in a system with a small number of active components? Unlike many suspensions of synthetic APs in Newtonian fluids, whose hydrodynamic interactions are usually negligible for their collective dynamics \cite{liebchen2019}, here the system is subjected to significant forces and torques mediated by the viscoelastic surrounding, spanning distances comparable or larger than the particle size. As demonstrated by the multiple examples described in previous sections, the strength of such forces and torques increases with increasing particle activity. Therefore, at small propulsion speed, the repulsive inter-particle interactions as well as the wall-induced potential are expected to be weak enough, thereby resulting in a liquid-like behavior of the active suspension, see Fig. \ref{fgr:manyparticles}(e). This is verified in Figs. \ref{fgr:transition}(a) and \ref{fgr:transition}(b), where we plot the mean square displacement $\langle \Delta r(t)^2\rangle = \langle [r(t+t_0) - r(t_0)]^2 \rangle$ of the radial particle position, $r$, and the mean square displacement $\langle \Delta \phi(t)^2\rangle = \langle [\phi(t+t_0) - \phi(t_0)]^2 \rangle - \langle \phi(t+t_0) - \phi(t_0) \rangle^2$ of the polar angle $\phi$, respectively. Here, a time average over $t_0$ is first computed for a given particle trajectory and then an ensemble average is performed over the $N = 5$ particles forming the active system. Indeed, we observe that at 5 $\mu$W$\mu$m$^{-2}$, both $\langle \Delta r(t)^2\rangle$ and $\langle \Delta \phi(t)^2\rangle$ exhibit clearly  diffusive behavior at sufficiently long times, thus resulting in liquid-like motion similar to that of passive colloidal suspensions under confinement below the transition to dynamical arrest~\cite{nugent2007,hunter2014}. In addition, in this regime we also find that that probability distribution $\bar{\rho}(r/R)$ of the radial position of an AP, where the bar represents an average over the total number $N = 5$ of particles, is rather broad, as plotted in Fig. \ref{fgr:transition}(c). 

Upon increasing the laser intensity $I$, the particles tend to  explore more rapidly the available space, but at the same time their higher activity instigates stronger torques and repulsive particle-particle interaction forces due to larger deformations of the viscoelastic medium. Consequently, APs repel each other and move away from the pore center. On the other hand, the forces induced by the curved wall push the particles to the pore center. If the particle activity is large enough, the corresponding propulsive forces are balanced by the viscoelasticity-mediated ones. In this manner, the whole system reaches a quasi-steady arrested state characterized by a ring-like structure with radial and polar order relative to the center of the confining pore, see Fig. \ref{fgr:manyparticles}(f). In such a case, we find that $\langle \Delta r(t)^2\rangle$ becomes subdiffusive due to the strong radial localization, whereas $\langle \Delta \phi(t)^2\rangle$ remains diffusive with a largely reduced slope, as shown in the insets of Figs. \ref{fgr:transition}(a) and \ref{fgr:transition}(b), respectively. Furthermore, the width of the radial distribution function $\bar{\rho}(r/R)$ decreases appreciably due to the radial localization, see Fig. \ref{fgr:transition}(c).
Note that such an ordered arrangement of the APs within the circular pore resembles qualitatively the formation of crystalline clusters in confined paramagnetic colloidal systems \cite{bubeck1999,schella2015}. In such cases, repulsive dipole-dipole interactions foster classical 2D atomic-like structures composed of concentric shells~\cite{peeters1995}, where the applied magnetic field plays a role similar to the particle activity in our multi-AP system.
For that kind of paramagnetic systems, the ordered clusters become more and more stable with increasing repulsion even in presence of thermal fluctuations, as torques are not exerted on individual particles.
Nevertheless, for a confined multi-AP system, increasing the activity also increases the strength of the induced viscoelastic torques. Then, if the particle activity is large enough, above a critical value $I_c$ of the laser intensity, the torques are able to destabilize the force balance and the structural order is smeared out, as shown in Figs. \ref{fgr:manyparticles}(g) and (h). Indeed, in Fig. \ref{fgr:transition}(a) we show that above $I_c$, $\langle \Delta r(t)^2\rangle$ displays a very rapid increase at short times due to the breakdown of the viscoelasticity-induced structural order, while a slower growth shows up at longer times due to the confinement created by the rigid cavity. On the other hand, $\langle \Delta \phi(t)^2\rangle$ remains diffusive even at $I > I_c$, where the corresponding slope shoots up with increasing activity, see Fig. Fig. \ref{fgr:transition}(b). The appearance of the re-entrant liquid-like phase translates into a marked broadening of the radial probability distribution of the particles, as found in Fig. \ref{fgr:transition}(c).

\begin{figure}
	\includegraphics[scale=0.65]{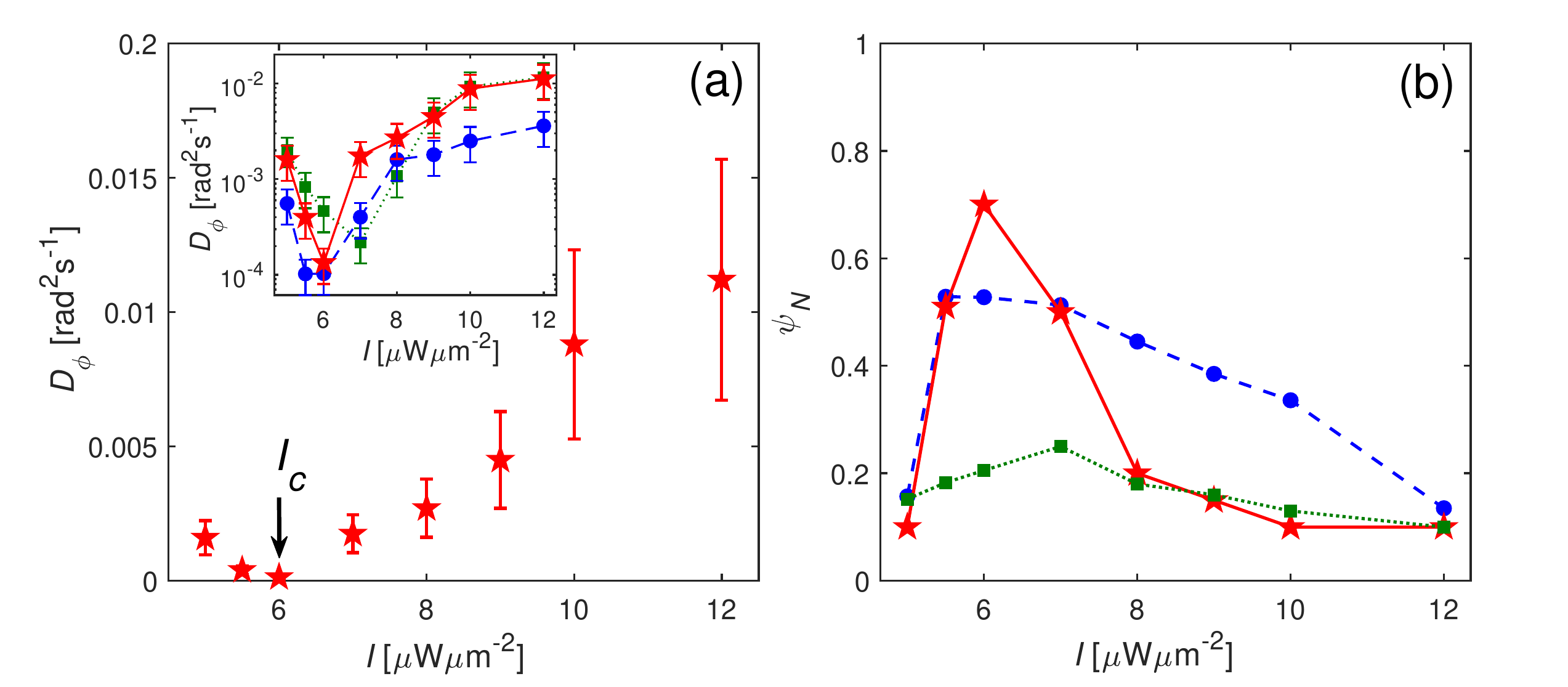}
	\caption{(a) Angular diffusion coefficient of self-propelled particles of diameter $2a=7.82$ $\mu$m as a function of intensity of the illumination for $N=5$. The arrow depicts the position of minimum angular diffusion at $I_c$. Inset: semilog-representation of $D_{\phi}$ for  $N=4$($\square$),~$N=5$ (\FiveStarOpen),~ $N=6$ ($\bigcirc$) in the viscoelastic fluid. (b) Angular order parameter as a function of light intensity for the each case mentioned above in (a) with the same symbols for the corresponding $N$.}
	\label{fig:angularorder}
\end{figure}

Since for sufficiently large time-scales $t$, $\langle \Delta \phi(t)^2\rangle$ is diffusive for all values of $I$, i.e.
\begin{equation}\label{eq:dphi}
    \langle \Delta \phi(t)^2\rangle =  2D_\mathrm{\phi} t, 
\end{equation}
we can characterize the order within the pore by means of the polar-angular diffusion coefficient $D_{\phi}$. We point out that $D_{\phi}$ results from the combined effect of the viscoelasticity-mediated interactions and thermal fluctuations in the fluid. In Fig. \ref{fig:angularorder}(a) we show that, for $N = 5$, $D_{\phi}$ has a strong non-monotonic dependence on $I$, with a pronounced drop of one order of magnitude at $I_c = 6 \mu$W/$\mu$m$^{-2}$ at which the particles self-organize into the ordered ring-like structure. For values $I > I_{c}$, $D_{\phi}$ sharply grows 2 orders of magnitude with respect to the minimum, thereby disclosing the appearance of the re-entrant liquid-like behavior of the active suspension. Besides, we can quantify the orientational order of the whole system by means of the parameter $\psi_{N}$ defined as
 \begin{equation}
\psi_{{N}}= \Bigg\langle \frac{1}{N^2}  \left| \sum_{j = 1}^N e^{iN\phi_j} \right|^2 \Bigg\rangle   
\end{equation}
where $\phi_{j}$ is the polar angle of the $j$-th particle ($j = 1,..., N$) as defined in Fig.~\ref{fgr:porenewtonianv1}(a), and the brackets denote the average over all frames. The order parameter is $\psi_{{N}} = 1$  for a perfect polygon shape of the corresponding $N$, while $\psi_{{N}} = 0$ for a fully disordered system. In  Fig.~\ref{fig:angularorder}(b), we show the dependence of $\psi_{{N}}$ on $I$ and find that it attains a maximum value $\psi_{{N}} \approx 0.7$ at $I_{c}$ for $N = 5$, thus confirming the emergence of high orientational order of the particles in the string-like structure. Such an orientational order is curtailed below $I_{c}$ because of the weakness of the repulsive interactions in the liquid-like phase, whereas above $I_{c}$ it also drops due to the increasing strength of the destabilizing torques in the re-entrant fluid phase.

We check that the proposed mechanism is consistent with supplementary observations of the multi-AP system. For instance, a slight increase (decrease) of the number of particles $N$ within a pore of fixed size will decrease (increase) the space available to them. For a fixed $I$, this will in turn change the strength of the forces and torques acting on each particle and the resulting correlations. Accordingly, it is expected that the values of $D_{\phi}$, $\psi_{N}$ and $I_c$ will depend on $N$. In fact, in the inset of Fig. \ref{fig:angularorder}(a) we demonstrate that the angular diffusion coefficient $D_{\phi}$ takes comparatively lower values for $N = 6$ (circles). This is because at a given activity (determined by $I$), the particles have less accessible space, thus resulting in closer inter-particle interactions. As a consequence, smaller particle activity is required to attain the ordered arrested state, which lowers the value of $I_{c}$, as shown in the inset of Fig. \ref{fig:angularorder}(a). In addition, it is expected that higher orientational order (compared to the case $N = 5$) persists for $I > I_{c}$. Indeed, this is in agreement with our results shown in Fig. \ref{fig:angularorder}(b) for which the values of $\psi_{N = 6}$ are larger than those of $\psi_{N = 5}$ at $I > 6 \mu$W$\mu$m$^{-2}$.
On the other hand, decreasing the number of particles to $N = 4$ leads to the opposite behavior. $I_{c}$ is shifted to higher values, as a larger particle activity is needed to induce collective order of the system at larger inter-particle distances, as seen in Fig. \ref{fig:angularorder}(a). Additionally, the order of the resulting structure for $N = 4$ at $I_{c}$ is in general less pronounced than for $N = 5$, as verified in Fig. \ref{fig:angularorder}(b) for the order parameter $\psi_{N}$.

\section{Conclusions}

 We have investigated the motion of active colloidal particles in a viscoelastic fluid under geometrical confinement. Unlike particle-wall and particle-particle interactions in Newtonian liquids, here we observe that they interact repulsively with the surface of a rigid wall. The repulsive interaction stems from the strain of the viscoelastic fluid between the particle and the wall and therefore strongly depends on its activity and the local curvature of the wall surface. Moreover, the spatial asymmetry imposed by the confinement induces significant hydrodynamic torques on the rotational motion of the particle, which would otherwise be governed by rotational diffusion. We have presented several examples of novel effects which originate from such viscoelasticity-mediated interactions. For example, a single AP exhibits very short residence times on planar wall surfaces, pronounced deflections by stationary obstacles, steering by arrays of obstacles and localization within circular pores. Remarkably, these effects also lead to the emergence of intriguing collective behavior of multi-AP systems under circular confinements. In particular, we uncover a transition from liquid-like behavior to a self-organized ordered state upon increasing their activity, whereas a further increase in the activity restores the liquid-like phase. Although the non-equilibrium phases resulting from such a transition have structural features similar to those in few-body passive colloidal systems under circular confinement, it is strikingly different in the sense that here they are dynamical and originate by the particle activity rather than external fields.
 Thus, our findings are expected to be of great importance for the understanding of self-organization of concentrated suspensions of microswimmers in confined geometries~\cite{wioland2013,lushi2014}. In such crowded systems, the interplay between confinement, specific swimming mechanisms, and the resulting hydrodynamic flows lead to a plethora of patterns that are difficult to predict from steric constraints and geometric considerations alone~\cite{tsang2015}.
 Furthermore, it would also be interesting to investigate the effect of other specific properties of circularly confined active systems, such as chirality~\cite{hoell2018}, and the combination of detailed swimming patterns with the geometrical curvature~\cite{ostapenko2018}, as they are also known to be responsible for strong localization of individual active particles even in Newtonian fluids.

\section*{Acknowledgements}
We thank M. Sahebdivani and J. Bois for their assistance during the early stages of this work. We acknowledge financial support of the Deutsche Forschungsgemeinschaft, BE 1788/10-1.

\appendix

\section{Determination of rheological parameters}\label{app:microrheology}

\begin{figure}
	\includegraphics[scale=0.65]{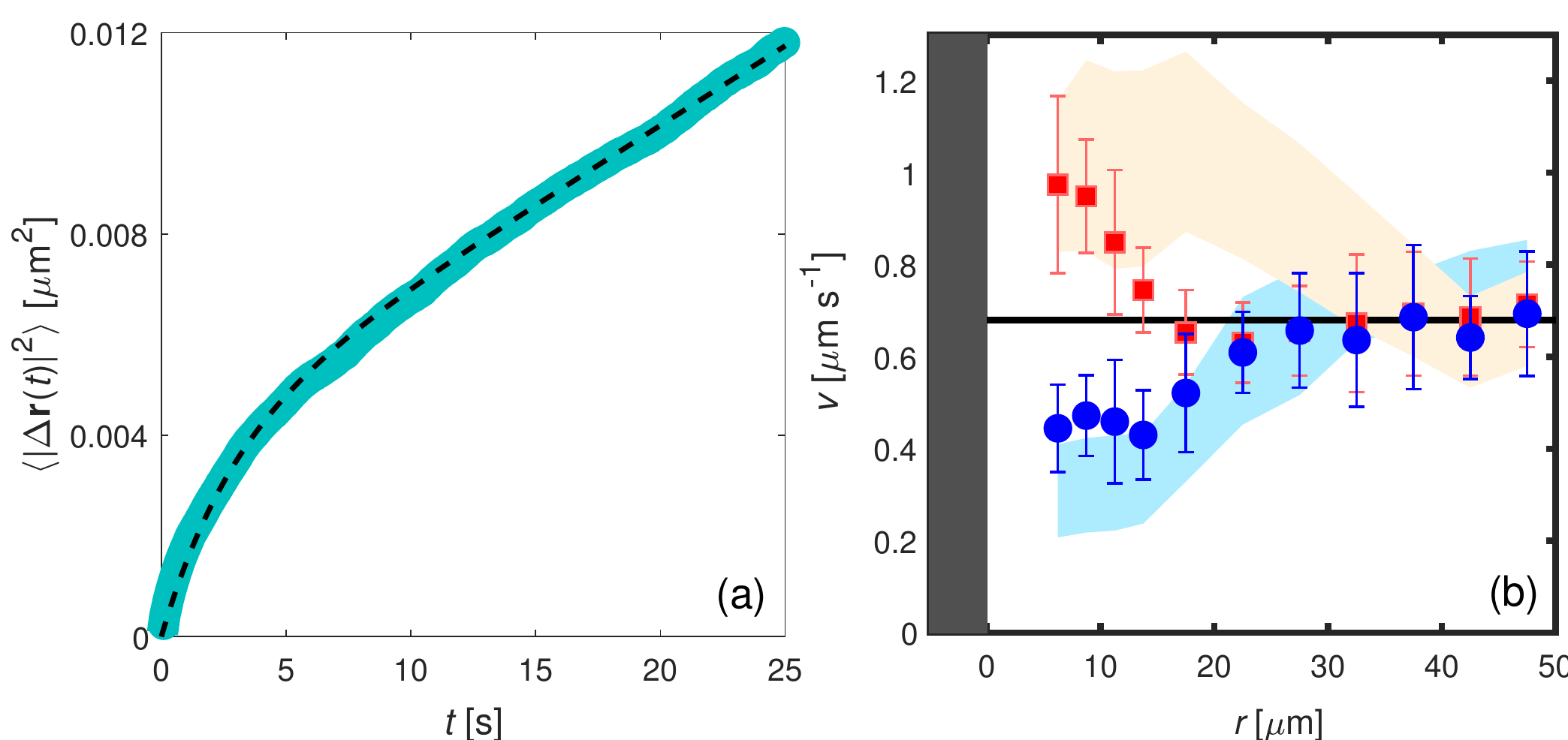}
	\caption{(a) Mean-square displacement of a passive particle (radius $a = 3.91\,\mu\mathrm{m}$) moving in the polymer solution at $0.05$ wt.\% and  $T = 298$~K (thick solid line). The dashed line represets a fit to Eq. (\ref{eq:msdpassive}), from which we find the rheological parameters $\eta_0 = 0.21$~Pa~s,  $\eta_{\infty} = 0.039$~Pa~s, and $G_0 = 9.9$~mPa. (b)  Particle speed $v$ as a function of the distance to the wall, $r$, in the viscoelastic fluid at  $0.03$ wt.\%  during the approach ($\circ$) and during the departure ($\square$). The horizontal solid line represents the particle speed very far from the wall, which is depicted as a vertical rectangle on the left ($r \le 0$) . The shaded areas correspond to the particle speed measured in the polymer solution at  $0.05$ wt.\%, see Fig. \ref{fgr:flatwall}(f).}
	\label{fig:appendix}
\end{figure}

In this section, we describe the method to characterize the linear viscoelasticity of the polymer solutions used in the experiments. We perform passive microrheology \cite{squires2010}, where the rheological parameters of the fluid under investigation are inferred from the free Brownian motion of an embedded colloidal particle. Because we are mainly interested in estimating the elastic force exerted by the strained fluid on a suspended particle, as well as rotational diffusion times in absence of active motion, we consider the simplest three-parameter rheological model for the relaxation modulus of the fluid
\begin{equation}\label{eq:relaxmodulus}
	G(t) = 2\eta_{\infty} \delta(t) + G_0\exp\left(-\frac{G_0}{\eta_0 - \eta_{\infty}} t\right),
\end{equation}
where $\eta_{\infty}$ is the solvent viscosity, $\delta(t)$ the Dirac-delta function,  $\eta_0$ the zero-shear viscosity, and $G_0$ the elastic modulus.  Eq.~(\ref{eq:relaxmodulus}) can be derived upon linearization of  the more general Stokes-Oldroyd-B model at vanishing Reynolds number \cite{paul2018}, which takes into account the instantaneous viscous response of the solvent, and the slower elastic response of the polymers with relaxation time $\tau = \frac{\eta_0 - \eta_{\infty}}{G_0}$ \cite{thomases2007}. Then, we assume a generalized overdamped Langevin equation for the position $\mathbf{r} = (x,y)$ of a spherical Brownian particle (radius $a$) embedded in the fluid with relaxation modulus (\ref{eq:relaxmodulus}) at temperature $T$ 
\begin{equation}\label{eq:GLE}
	0 = -6\pi a \int_{- \infty}^t \mathrm{d}t'  G(t - t') \dot{\mathbf{r}}(t') + \zeta(t),
\end{equation}
where the convolution accounts for the response of the frictional force acting on the particle to the viscoelastic surrounding, whereas  $\zeta$ represents a thermal colored noise of zero mean, $\langle  \zeta(t) \rangle = 0$, and autocorrelation $\langle \zeta(t) \zeta(s) \rangle = 6\pi a k_B T G(|t -s|)$. From Equations (\ref{eq:relaxmodulus}) and (\ref{eq:GLE}), a straightforward calculation leads to an expression for the 2D mean-square displacement of the particle position 
\begin{equation}\label{eq:msdpassive}
	\langle |\Delta \mathbf{r}(t)|^2 \rangle = \frac{4 k_B T}{6 \pi a \eta_0} \left \{ t + \frac{(\eta_0 - \eta_{\infty})^2}{G_0\eta_0}\left[  1 - \exp \left( -\frac{t}{\lambda}\right) \right] \right \},
\end{equation}
where $\lambda = \left( 1 - \frac{\eta_{\infty}}{\eta_0} \right) \frac{\eta_{\infty}}{G_0}$. For $t \ll \lambda$ and $t \gg \lambda$, Equation (\ref{eq:msdpassive}) correctly reproduces short-time and long-time diffusive behaviors, which are expected as a results of the two distinct friction coefficients $6\pi a \eta_{\infty}$ and $6\pi a \eta_0$, respectively.
In Fig.~\ref{fig:appendix}(a) we demonstrates that Eq.~(\ref{eq:msdpassive}) describes well the mean-square displacement of a passive particle of radius $a = 3.91\,\mu\mathrm{m}$ moving in the polymer solution at a concentration 0.05 wt.\% of PAAm and $T = 298$~K. From the fit of the experimental data to Equation (\ref{eq:msdpassive}), we find the values $\eta_0 = 0.210$~Pa~s, $\eta_{\infty} = 0.039$~Pa~s, and $G_0 = 9.9$~mPa reported in the main text.

\section{Role of viscoelasticity}\label{app:viscoelasticity}
Here, we present some results of the motion of an AP near a planar wall at lower PAAm concentration, which demonstrate that the smaller the elasticity of the fluid relative to its zero-shear viscosity, quantified by the ratio $\frac{G_0 \tau}{\eta_0} = 1 - \frac{\eta_{\infty}}{\eta_0}$, the weaker the effect of the solid wall surface on the active motion. 
For instance, at  $0.03$ wt.\%, the rheological parameters are $G_0 = 1.3$~mPa, $\eta_0 = 0.053$~Pa~s, and $\eta_{\infty}= 0.039$~Pa~s, thus leading to $1 - \frac{\eta_{\infty}}{\eta_0} = 0.26$, which is smaller than the value $1 - \frac{\eta_{\infty}}{\eta_0} = 0.81$ at $0.05$ wt.\%.
In Fig.~\ref{fig:appendix}(b) we show that the effect of the wall on the particle speed, $v$, is less pronounced when moving in the polymer solution at lower concentration ($0.03$ wt.\%). First, when the particle gets close to the wall, the speed begins to decreases at a distance $r \approx 5a$ with respect to the value far away from the wall. Such a decrease is less pronounced than that observed in the fluid at $0.05$ wt.\%. Then, the particle remains almost in contact with the wall ($r \approx a$), where the resulting  residence time is approximately 9~min. This value is significantly larger than that at higher PAAm concentration (1~min at $0.05$ wt.\%, see main text), but still orders of magnitude shorter than the rotational diffusion time given by the Stokes-Einstein relation $D_r^{-1} =\frac{8 \pi \eta_0 a^3} {k_BT}\approx 5$~h. After that, the particle is able to reorient  and to be detached from the wall, with a speed which is larger than that in the bulk due to the sudden stress release of the fluid. Finally, $v$ reaches the bulk value at a distance $r \approx 3a$, which is also smaller than the corresponding distance in the fluid at $0.05$ wt.\%.

\end{document}